\documentclass[twocolumn,showpacs,preprintnumbers,amsmath,amssymb]{revtex4}
\usepackage{graphicx}
\usepackage{epstopdf}
\usepackage{dcolumn}
\usepackage{bm}
\newcommand{\rr}{{\bf r}}
\newcommand{\vv}{{\bf v}}

\newcommand{\BE}{\begin{equation}}
\newcommand{\EE}{\end{equation}}
\newcommand{\be}{\begin{equation}}
\newcommand{\ee}{\end{equation}}
\begin{document}

\preprint{Phys. Rev. E.{\bf 73}, 040901 (2006)}

\title{Anomalously Slow Domain Growth in 
Fluid Membranes with Asymmetric Transbilayer Lipid Distribution}

\author{Mohamed Laradji$^{1,3}$ and P. B. Sunil Kumar$^{2,3}$}
\affiliation {$^1$Department of Physics, The University of Memphis, 
Memphis, TN 38152 \\
$^2$Department of Physics, Indian Institute of Technology Madras,
Chennai 600036, India\\
$^3$MEMPHYS--Center for Biomembrane Physics, University of Southern Denmark, DK-5230, Denmark}

\begin{abstract}
The effect of asymmetry in the transbilayer lipid distribution on 
the dynamics of phase separation in fluid vesicles 
is investigated numerically for the first time. This asymmetry is shown 
to set a spontaneous curvature for the domains that alter the morphology and dynamics considerably.
For moderate tension, the domains are capped and the spontaneous curvature leads to anomalously slow dynamics, as compared to the case of symmetric bilayers. In contrast, in the limiting cases of high and low tensions, the 
dynamics proceeds towards full phase separation.
\end{abstract}


\maketitle

Asymmetric distribution of lipids in the two leaflets of the plasma membrane is ubiquitous to many eukaryotic cells.
Most of phosphatidylserine and phosphatidylethanolamine are located in the cytoplasmic leaflet, while 
sphingomyeline and phosphatidylcholine are predominantly in the outer leaflet~\cite{warren87}. This asymmetry,
maintained by the cell through many active and passive processes,
plays an important role in the lateral and trans-membrane compositional and morphological
organizations in the nanometer scale. A very good example for such organization  is the 
 nanoscale domains, referred to as rafts. These domains are 
 believed to be liquid-ordered regions, mainly composed of sphingomeylin, which 
is a saturated lipid, and cholesterol~\cite{kusumi04}.  In spite of the wealth of experimental studies on lipid rafts, the mechanisms leading to their formation and their stability remain under intense debate. These issues are complicated by the  presence of many components and processes in biomembranes.

With the aim to achieve the understanding of the physical properties of biomembranes, many experimental~\cite{keller05}  and theoretical~\cite{lipowsky95,laradji04} investigations have been carried out on relatively simple model lipid membranes. 
To the best of our knowledge, in all these studies, membranes have the same lipid composition in both leaflets of the bilayer. The natural next step in complexity towards the understanding of biomembranes is 
to consider membranes with different lipid composition in the two leaflets. 
An important question that arises is then: What role does this transbilayer  asymmetry play on the domain structure of lipid bilayers? In particular, will this asymmetry result in a finite size of these domains?
The purpose of the present paper is to address these questions  using large scale dissipative particle dynamics (DPD) simulations.

Recent experiments clearly demonstrated that in multicomponent 
lipid vesicles there is a strong registration of the domains in the two leaflets. That is 
the type of lipids in the two leaflets are locally the same.    
This implies that, if there is a transbilayer asymmetry in the average lipid composition,
these domains in register in the two leaflets will have to be of different areas. In order to minimize the interaction energy between unlike lipids the domains have to curve as shown schematically in Fig.~1. The area difference between the domains in the two leaflets should then give rise to a spontaneous curvature. 
This spontaneous curvature is only a function of the average compositions 
in the two leaflets, $\phi_{out}$ and $\phi_{in}$ defined as the number of A-lipids in
the outer and inner leaflets, respectively, divided by the total number of lipids in the same leaflet
\begin{equation}
   c_0=\left(\frac{2}{\epsilon}\right)
[(\phi_{out}/\phi_{in})^{1/2}-1]  \, / \, [(\phi_{out}/\phi_{in})^{1/2}+1], \label{eq:spon}
\end{equation}
where, $\epsilon$ is the bilayer thickness.

\begin{figure}
\includegraphics[scale=0.3]{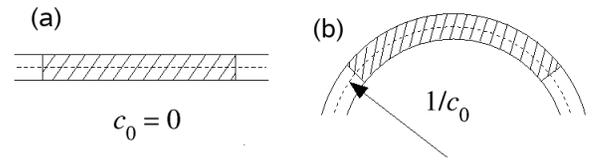}
\caption{(a) The configuration that minimizes the interaction energy between the lipids of a bilayer when the domains in the two layers are of the same area  and  (b) when 
the areas of the domains are different. The spontaneous curvature resulting from this 
area difference can be estimated by Eq.~(1).}
\end{figure}

Mean field calculations of multicomponent monolayer membranes, with  curvature coupled to the local concentration through a spontaneous curvature,  predict that 
when one of the components has a preferred curvature, 
the equilibrium state is that of caps or stripes.
These calculations, considering an ordered patterns of domains, have been 
carried out both in two- ~\cite{andelman92,gozdz01,harden05,hansen98,kumar99} 
and three-component~\cite{kumar99} monolayer models  for  lipid 
membranes. An ordered array of buds has also been proposed 
in the strong segregation limit~\cite{gozdz01,harden05}. These calculations are performed on an infinite planar membrane. An extension of these calculations to the case of finite closed vesicles are difficult and 
have not yet been reported. It is important to note that the Hamiltonian used in these monolayer models are based on a 
local bilinear coupling of the form, $\int_{\cal A} \phi H$, where $H$ is the local mean curvature and $\phi$ 
is the local composition field. One way to obtain 
these models from  two-component bilayer models amounts to having a local mismatch between types of lipids  
in the outer and inner leaflets~\cite{hansen98,kumar99}, and therefore
absence of domain register. Instead, a strong register  reported in experiments suggests that it is more reasonable  to expect a spontaneous curvature generated by the difference in area between the domains that are 
in register.

Motivated by the argument above, we carried out a systematic DPD simulation,
of self-assembled lipid bilayer vesicles with different lipid compositions in the two leaflets of the bilayer.  
Within the DPD approach~\cite{hoogerbrugge92} a number of atoms or molecules are coarse-grained 
in order to form a fluid element, thereafter called a dpd particle. Specifically, we have in our case three types of beads,
corresponding to a water-like bead, labeled {\em w}, a hydrophilic bead labeled {\em h}, representing a lipid head group, 
and a hydrophobic bead, labeled {\em t}, representing a group of ${\rm CH}_2$'s in the tail group~\cite{yamamoto03,ilya05}.
The model parameters are selected such that the membrane is impermeable to the solvent thus allowing, in the case of a 
vesicle, to investigate the effect of conservation of inner volume.  As in our previous study a lipid particle is
modeled by a fully flexible linear amphiphilic chain, constructed from one hydrophilic {\em h}-particle, 
connected to three consecutive hydrophobic {\em t}-particles~\cite{laradji04,laradji05}. There are two types of lipid particles, corresponding to $A$
and $B$-lipids. The heads and tail dpd particles of an $\alpha$-lipid, with $\alpha= A$ or $B$, 
are labeled $h_\alpha$ and $t_\alpha$, respectively.
The position and velocity of each dpd-particle are denoted by $\rr_i$ and $\vv_i$, respectively. Within the DPD approach, 
all particles are soft beads that interact with each other through three pairwise forces corresponding to a conservative
force, {\tiny $F^{(C)}_{ij}$}, a random force, {\tiny  $F^{(R)}_{ij}$}, and a dissipative force, {\tiny  $F^{(D)}_{ij}$}. The conservative force
between dpd particles, $i$ and $j$ is given by  {\tiny $ F^{(C)}_{ij}$}$=a_{ij} \omega(r_{ij}){\bf n}_{ij}$ where 
$\rr_{ij}=\rr_i-\rr_j$ and ${\bf n}_{ij}=\rr_{ij}/r_{ij}$. 
Since all dpd particles are soft, we choose $\omega(r)=1-r/r_c$ for $r\le r_c$, with $r_c$ is the cutoff of interaction 
and is used to set a length scale in the simulations.
$\omega(r)=0$ for $r> r_c$. 
The integrity of a lipid chain is ensured through a simple harmonic interaction between
consecutive monomers, {\tiny $ {\bf F}^{(S)}_{ij}$}$=-C( 1-|{\bf r}_{ij}|/b){\bf n}_{ij}$, where the spring constant $C=100\epsilon$ and
the length $b=0.45 r_c$. The dissipative force, describing friction between neighboring particles, is given by
{\tiny ${\bf F}^{(D)}_{ij}$}$= -\gamma \omega^2(|{\bf r}_{ij}|)({\bf n}_{ij}\cdot{\bf v}_{ij}){\bf n}_{ij}$, where $\vv_{ij}=
\vv_i-\vv_j$. Finally, the random force is given by, 
{\tiny ${\bf F}^{(R)}_{ij}$}$ = -\sigma \omega(|{\bf r}_{ij}|)\zeta_{ij}(\Delta t)^{-1/2}{\bf n}_{ij}$
where $\zeta_{ij}$ is a random noise with zero mean and unit variance, and 
$\Delta t$ is the time step of the simulation. The fluctuation-dissipation theorem requires
that $\gamma_{ij}=\sigma_{ij}^2/2k_{\rm B}T$. We used a fixed value for the parameter $\sigma_{ij}=\sigma$
for all pairs of dpd-particles. The head-head and tail-tail interactions between unlike lipids are given by 
$a_{h_A,h_B}=50 \epsilon$ and $a_{t_A,t_B}=50 \epsilon$ respectively with $\epsilon$ setting a scale for energy. 
All other parameters used in the present study are the same as in references~\cite{laradji04,laradji05}. 
In our simulations, a vesicle is composed
of 16 000 lipid particles, embedded in a fluid consisting of 1 472 000 
solvent particles with density $\rho=3r_c^{-3}$, and the time step, $\Delta t=0.05 \tau$ 
~\cite{Jakobsen05} with $\tau=(m r_c^2/\epsilon)^{1/2}$ and $m$ is the mass of a single dpd particle.

The interactions above ensure that the membrane is impermeable and that flip-flop events are extremely rare. 
The tail-tail interaction between two like lipids is
less repulsive than that between two unlike lipids. 
Different tail-tail interactions are needed in order to mimic the fact that the hydrophobic
region in domains which are rich in the saturated lipid and cholesterol is conformationally
different from domains rich in unsaturated lipids.
The phase separation is initiated through randomly labeling a fraction of all lipid particles as types {\em A}
or {\em B}, such that the outer and inner leaflets have compositions $\phi_{out}$ and $\phi_{in}$, respectively. 
The coarsening dynamics is then monitored through the computation of the cluster sizes of the 
minority lipid and the net interfacial length between the segregated domains. By simultaneously
monitoring both interfacial length and the average domain area, we are able to investigate the physical mechanism
via which domain growth proceeds.

\begin{figure}
\includegraphics[scale=0.7]{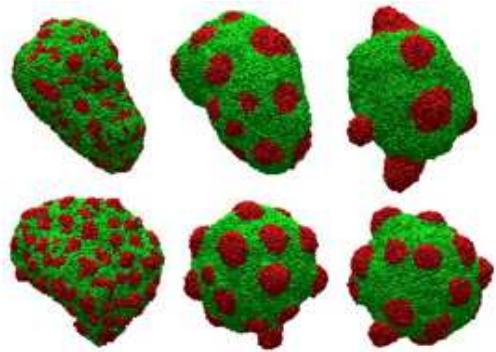}
\caption{Snapshot sequence of System II  (bottom row) as compared to 
that of a symmetric vesicle having the same area to volume ratio (top row). 
Snapshots from left 
to right correspond to $t=100$, $2000$ and $4000\tau$, respectively.}
\label{snaps_vesicle_40_20}
\end{figure}

Here, we present results for the case of 
$(\phi_{out},\phi_{in})=(0.4,0.2)$ with three different area-to-volume ratios, 
corresponding to the case of the number of solvent particles in the vesicle's 
core, $N_s=138000$ (System I), $N_s=112400$ (System II), and $N_s=72300$
(System III). In the remaining of this letter, the symmetric case, refers to a vesicle with
$(\phi_{out},\phi_{in})=(0.3,0.3)$ and an area-to-volume ratio equal to that
in System II. In Fig.~2, a sequence of snapshots in the case of system II, with $(\phi_{out},\phi_{in})=(0.4,0.2)$ are compared to that
of a vesicle with symmetric transbilayer composition corresponding to $(\phi_{out},\phi_{in})=(0.3,0.3)$.
This figure clearly shows that the onset of domain capping is shifted to earlier times by the transbilayer asymmetry in
the lipid composition, as opposed to the case of a symmetric vesicle. 
This capping result from the register between domains in the outer and inner
 leaflets and
mismatch between their areas (see Eq.~(\ref{eq:spon})). 

\begin{figure}
\includegraphics[scale=0.25]{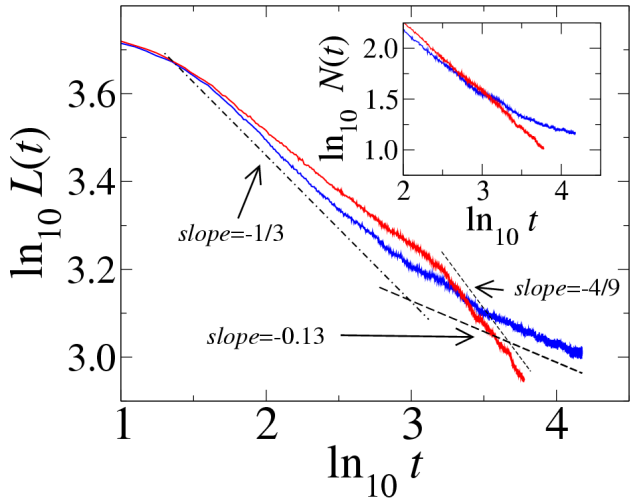}
\vskip.5cm
\caption{Net interface length as a function of time. The data with the slope $-4/9$ corresponds to System II and that with slope -0.13 corresponds to the symmetric vesicle. Inset shows the net number of domains as a function of time for System II (top curve at late times) 
and for the symmetric vesicle (bottom curve).}
\label{interface}
\end{figure}

At late times the coarsening dynamics is slower in the asymmetric case than in the
symmetric case as is evident from Fig.~2. This is substantiated by the time dependence of the net interface lengths of the two
systems shown in Fig.~3. 
This slowing down is not observed in the symmetric case even during late times, which   the expected fast growth law, $L\sim t^{-4/9}$, resulting from coalescence of 
caps~\cite{laradji04}.
Note that while the curvature of domains in the symmetric case is set by the competition between line tension and bending
modulus~\cite{lipowsky95}, in the asymmetric case it is induced by the 
spontaneous curvature  resulting from the asymmetry in the compositions of the two leaflets, 
and is set at early times (see Fig.~2).
The late times slowing down  shown in Fig.~3,
 must be the result of this spontaneous curvature. It is
interesting to note that  during the late stages, the interfacial length 
has a very small growth exponent ($\sim -0.13$), indicative of a logarithmic growth. This is normally attributed to microphase separation, ubiquitous to many other soft materials~\cite{laradji94}. This non-algebraic slow dynamics is 
the result of the competition between interfacial tension, which
is the driving force of the phase separation, and an effective long-range repulsive interaction. In the present case this repulsive interaction 
should result from the combined effect of spontaneous curvature and lateral tension. 
Microphase separation in multicomponent lipid bilayers with spontaneous curvature has been predicted by
mean field theories both in the weak and strong segregation limits at low values of lateral tension
~\cite{andelman92,hansen98,kumar99,gozdz01,harden05}. 
We should note that these theories assume that domains, in the microphase separation regime, are 
organized into a regular lattice. However thermal fluctuations could easily destroy the long-range order of
these domains. Indeed, in our simulation, we see a melt of caps instead of a regular lattice.

\begin{figure}
\includegraphics[scale=0.6]{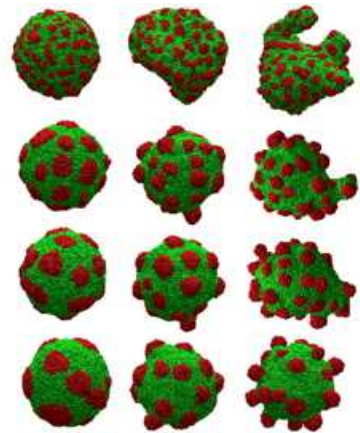}
\caption{Snapshot sequences for the three area to volume ratios considered in the present
study. Rows  from top to bottom represent the Systems-I, II, and III,
respectively. Snapshots from left to right, in 
each row, correspond to times $t=100$, 1000, 2000, and $5000\tau$, respectively.}
\label{snapshots-I}
\end{figure}

\begin{figure}
\includegraphics[scale=0.25]{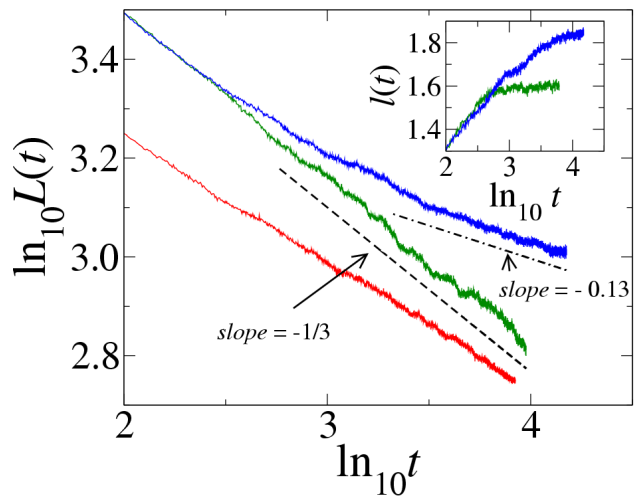}
\caption{Net interface length versus time. Curves from bottom to top
corresponds to Systems I (bottom), III (middle), and II (top), respectively. 
The data line for System I (bottom) hass shifted downwards for clarity. Straight lines are guides to the
eye.
In the inset, the average interface length per domain is shown versus time in Systems II (top curve), and III (bottom curve).}
\label{interface_length2}
\end{figure}

Surface tension is expected to play an important role on the phase separation of multicomponent lipid membranes. 
In the case of closed and impermeable vesicles with conserved number of lipids, the role of surface tension 
is effectively played by the area-to-volume ratio. Experimentally, surface tension is  
controlled by an osmotic pressure between the inner and outer environments of the vesicle.
In Fig.~4, snapshots of vesicles with asymmetric composition 
are shown for three different values of the area-to-volume ratio, corresponding to Systems I, II and III. 
At low tension (System III), the large amount of excess area allows domains 
to acquire a bud shape, such that interfaces of the corresponding domains in each leaflet are in register. 
Due to similar reasons, at intermediate tensions (System II) domains acquire a cap shape 
since the excess area available in this case is less than that in System III. 
At high tensions (System I), the lack of excess area prevents domains from curving, leading 
them to remain nearly flat throughout the phase separation process. 

The net interface lengths of the domain structure versus time are shown in Fig.~5 for the three 
different tensions.  This figure shows that during early times of the phase separation process, 
i.e. for $t\lesssim 250$ the dynamics is independent of tension.
As demonstrated by the early time configurations in Fig.~4, this is due to the fact that during this regime,  
domains are flat, and the composition dynamics is decoupled from the membrane curvature dynamics. 
At late stages, the dynamics in the three systems become noticeably different. 
We notice in particular that the dynamics is very slow in the case of intermediate tension (System II), 
as discussed above. In contrast, domain coarsening is faster and algebraic for both high (System I) and low (System III) 
tension, implying an approach towards full phase separation. A growth law, $L(t)\sim t^{-1/3}$ in System I, 
is the result of coalescence of flat domains as is shown earlier in Refs.~\cite{laradji04,laradji05}.
Our results for cases of intermediate and high tensions are in agreement with recent theoretical
predictions~\cite{harden05}. However, for the low tension case, we see growth dynamics towards
full phase separation.  The equilibrium state here will be a completely 
phase separated vesicle~\cite{julicher95}.

The diffusion coefficient of a single bud should scale as  $D\sim 1/a^{1/2}$, where $a$
is the area of a single bud. If domain coarsening at late stages of system III is mediated by 
coalescence of spherical buds, with
a fixed neck radius $l$, diffusing on the surface of the vesicle (the length scale here being set by the ratio of  difference in area occupied by the B phase to the bilayer thickness) then the net interface length,
$L(t)=N(t)l\sim t^{-2/3}$, where  $N$ is the total number of buds on the vesicle
~\cite{kumar01}. 
This growth law is much faster than what is shown in Fig.~5.  $l(t)=L(t)/N$ for Systems II and III
plotted in the inset of Fig.~5. This figure shows that the interfacial length per bud,  and hence the neck radius, in system III reaches saturation at
about $t\approx 250\tau$. A close examination of snapshots at late times clearly indicate that domain growth in system III is mediated
by coalescence. Therefore, in order to reproduce the observed time dependence, $L(t)\sim t^{-1/3}$,
the buds must move with an effective diffusion coefficient that scales as $1/a^2$. Since this is 
unrealistic there must be another process opposing 
the necks of two buds from merging. The details of this process is currently under investigation.

In conclusion, we examine the effect of transbilayer asymmetry in the lipid composition
on phase separation. We found that at intermediate 
tension, the asymmetry leads to anomalously slow coarsening of caps, in agreement with recent mean field calculations.
At low or vanishingly small lateral tension, where the domain structure is
that of buds,  and at  high tension, where the  domains are flat, we found algebraic domain growth leading to full phase separation. 
It would be very useful to explicitly correlate the nature of domain
growth, with asymmetric transbilayer 
lipid distribution, with the surface tension of the membrane. We plan to
perform the calculations of the surface tension in the various regimes from
the lateral stress profiles using the approach similar to that in Refs.
~\cite{ilya05,shillcock02}.

M.L. thanks the Petroleum Research Fund and P.B.S.K. thanks the DST, India, for
support.  MEMPHYS is supported by the Danish National Research Foundation. 
Parts of the simulation were carried at the Danish Center for Scientific Computing.

\end{document}